\documentclass[5p,preprint,longtitle,times,twocolumn]{elsarticle} 

\usepackage{lineno}
\usepackage{hyperref}
\usepackage{amsmath,amsfonts,amsthm,amsbsy}
\usepackage{amssymb}
\usepackage{units}

\modulolinenumbers[5]

\journal{Journal of \LaTeX\ Templates}

\begin{document}

\begin{frontmatter}

\title{A Cryogenic Supersonic Jet Target for Electron Scattering Experiments at MAGIX@MESA and MAMI}

\author[ikp]{S. Grieser\corref{cor1}}
\ead{s$\_$grie06@uni-muenster.de}
\cortext[cor1]{Corresponding author}

\author[ikp]{D. Bonaventura}
\author[ikp]{P. Brand}
\author[ikp]{C. Hargens}
\author[ikp]{B. Hetz}
\author[ikp]{L. Le\ss{}mann}
\author[ikp]{C. Westph\"alinger}
\author[ikp]{A. Khoukaz}
\ead[url]{www.uni-muenster.de/Physik.KP/AGKhoukaz/index.html}
\address[ikp]{Institut f\"ur Kernpysik, Wilhelm-Klemm-Str. 9, 48149  M\"unster, Germany}





\begin{abstract}
High-performance cluster-jet targets are ideally suited and applied since years in hadron and laser plasma physics. Therefore, the forthcoming MAGIX experiment at the future energy recovering electron accelerator MESA will use a cluster-jet target to perform high precision measurements on electron scattering experiments, i.e., determination of the proton radius. For this purpose, a cluster-jet target was designed, built up and set successfully into operation at the University of M\"unster considering the requirements of the experimental setup of MAGIX. The details on these requirements, calculations to their realization, e.g., on the nozzle geometry and stagnation conditions of the target gas, their technical implementation and the features of the target which make the target a powerful state-of-the-art target, are highlighted in this publication. Furthermore, the measured and analysed jet beam characteristics from this target using a Mach Zehnder interferometer are presented and discussed. These are of highest interest for the final design of the complete experimental setup of MAGIX. Moreover, first measurements from commissioning beam times performed with the target installed at the already running MAinzer MIkrotron will be presented.  
\end{abstract}


\end{frontmatter}


\section{Introduction}

One essential constituent of the upcoming MAGIX (MesA Gas Internal target eXperiment) experiment at MESA (Mainz Energy-recovering Superconducting Accelerator) will be a
high-performance cluster-jet target.
Cluster-jet targets are highly suited and established as targets for hadron and laser plasma physics \cite{Brauksiepe:1996ii,Dombrowski:1997ba, Khoukaz1999,Taschner:2011ew,Jinno:2018}. They provide a continuous stream of target material by the expansion of (pre-cooled) gases through a fine Laval nozzle. Such targets achieve a high and constant beam thickness, which is at the same time adjustable over several orders of magnitude by changing the stagnation conditions of the gas in front of the nozzle. \\
Cluster-jet targets were already applied as internal targets, e.g., for the COSY-11 or ANKE (Apparatus for Studies of Nucleon and Kaon Ejectiles) experiments at the COSY (COoler SYnchrotron) accelerator in J\"ulich (Germany) \cite{Dombrowski:1997ba,Barsov:2001xj,Wilkin:2016mfn}. These targets realised hydrogen target thicknesses of more than $\unit[10^{15}]{atoms/cm^{2}}$ in a distance of $\unit[65]{cm}$ from the nozzle \cite{Dombrowski:1997ba,Stein:2008ga}. The future $\overline{\text{P}}$ANDA (anti-Proton ANnihilations at DArmstadt) experiment at the FAIR (Facility for Antiproton and Ion Research) facility will use a cluster-jet target as the first operating target \cite{ TargetTDR:2012}. The $\overline{\text{P}}$ANDA hydrogen cluster-jet target prototype was designed and built up at the University of M\"unster and is routinely in operation since years and achieves a thickness of more than $\unit[10^{15}]{atoms/cm^{2}}$ at distance of $\unit[2.1]{m}$ from the nozzle \cite{Khoukaz:2013nyd}. Moreover, the final $\overline{\text{P}}$ANDA cluster-jet target is also built up and already set successfully into operation and will be installed at COSY in 2018 for detailed beam-target interaction studies \cite{PandaProposal}. The realisation of the high target thickness at these large distances in combination with a well defined cluster beam to minimize the influence on the vacuum condition of the accelerator beam line is very important. Therefore, a set of orifices is applied to separate the cluster beam from the residual gas and to size and shape the beam \cite{PhdKoehler}. \\
While hadron physics experiments at internal synchrotron target points typically require target thicknesses of up to $\unit[10^{15}]{atoms/cm^{2}}$, experiments like the CryoFlash experiment to study cluster-laser interactions performed by the University of M\"unster in cooperation with the University of D\"usseldorf \cite{Willi}, and the planned MAGIX experiment at MESA aim for significant higher target thicknesses, i.e. $\rho_\text{areal} > \unit[ 10^{18}]{atoms/cm^2}$. Therefore, the interaction between the laser or electron beam with the cluster beam will be performed directly behind the nozzle without any further use of beam skimmers.  \\
The MAGIX experiment will consist of a jet target and two identical spectrometers equipped with GEM detectors (Gas Electron Multiplier) as focal plain detectors, to measure the angle and momentum of the scattered particles and will be located in the energy recovering sector of MESA \cite{SCaiazza}.
There an electron beam with a beam current of  $I = \unit[10]{mA}$ will be provided \cite{Hug:2017ypc}, corresponding to a possible event rate of $\dot N_\text{int} = I/e \approx  \unit[ 6.2 \times 10^{16}]{s^{-1}}$. The design luminosity for the MAGIX experiment of $L = \unit[10^{35}]{cm^{-2}s^{-1}}$ corresponds via 
	\begin{equation}
	L = \dot N_\text{int}  \cdot \rho_\text{areal}
	\end{equation}
to an areal target thickness of $\rho_\text{areal} \approx \unit[1.6 \times 10^{18}]{atoms/cm^2}$ for the target. The target presented here is designed to achieve even higher target thicknesses to be more flexible to the future experimental program planned at MAGIX. Moreover, to minimize the background reaction rates, a jet target has the advantage that the target material is directly injected into the scattering chamber, thus there is no need for foils or windows between target and accelerator beam. Another further measure to minimize the background is the optimization of the vacuum conditions in the scattering chamber. Therefore, a catcher system opposite to the nozzle will be installed to guide the cluster beam directly after the interaction point with the accelerator beam out of the scattering chamber. Depending on the experimental program at MAGIX, the target offers the possibility to be operated with different gases as target material. Besides that, the nozzle geometry can be adapted to the requests for the desired gas, target thickness, size and shape of the cluster beam. The jet target was designed, built up and set successfully into operation at the University of M\"unster \cite{PhdGrieser, HeraeusGrieser, BAHargens}.\\
The MAGIX experiment will use electron scattering for precision measurements of fundamental constants, such as the proton radius or the astrophysical S-Factor \cite{Merkel:2016bfj}. The precise measurement of the proton charge radius is essential to solve the existing proton radius puzzle. Measurements with muonic hydrogen resulted in a radius of $r_p = \unit[ \left( 0.84184 \pm 0.00067 \right) ]{fm}$ \cite{Antognini:2011cxa}. In comparison to the CODATA value of $r_p = \unit[ \left( 0.8751 \pm 0.0061 \right) ]{fm}$ \cite{CODATA} which is dominated by studies on laser spectroscopy with hydrogen, this value is $ \unit[ 5 ]{\sigma}$ lower, but ten times more precise. A following measurement using electron-proton scattering at MAMI yielded a charge radius of $r_p = \unit[ \left( 0.879 \pm 0.008 \right) ]{fm}$ \cite{Bernauer:2010wm}, which enlarged the discrepancy up to $ \unit[ 7 ]{\sigma}$. A new approach to measure the proton charge radius applies the initial state radiation technique. First results lead to a radius of $r_p = \unit[ \left( 0.810 \pm 0.082 \right) ]{fm}$ \cite{Mihovilovic:2016rkr}. Due to the relatively large uncertainty, new measurements using this technique with the jet target presented here for MAGIX will be performed to confirm the proton radius \cite{HMerkel}. \\
While the MAGIX@MESA facility with its energy recovering accelerator is still in preparation, the target is already available for optimisation and preparatory studies on the beam-target interaction and is planned to be used for first electron scattering experiments at a ''conventional'' accelerator. For this purpose since August 2017, the target is installed at the MAMI (MAinzer MIkrotron) electron accelerator complex in Mainz (Germany) to perform, e.g., additional measurements on the proton radius puzzle.\\

      
\section{Experimental Setup}
\begin{figure}[!ht]
    \centering
    \includegraphics[trim = 0mm 0mm 0mm 0mm, clip, width=0.48\textwidth]{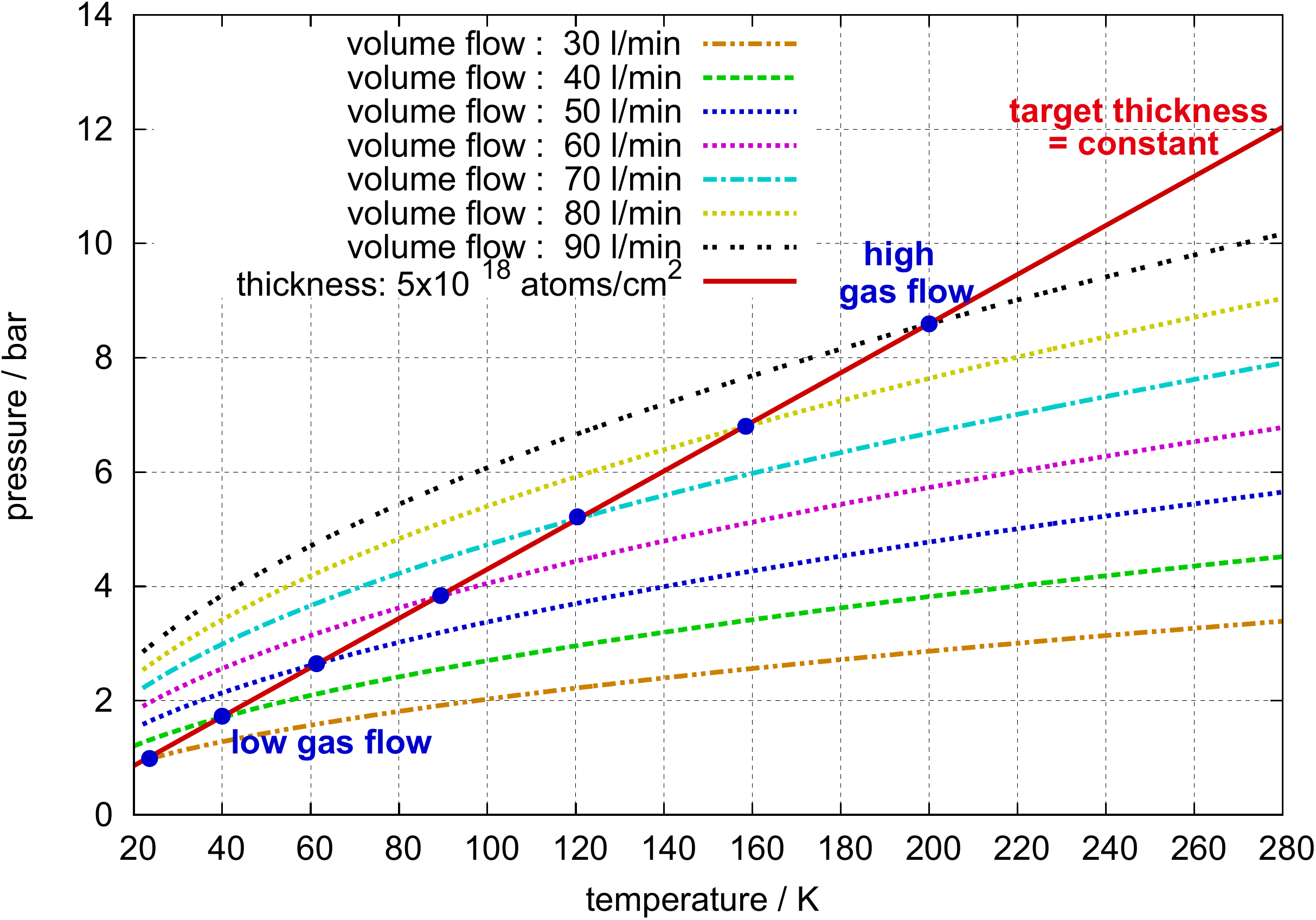}
    
    \caption{Coloured curves indicate different constant volume flows through the nozzle achievable with various temperature and pressure combinations. Exemplary pressure-temperature pairs with a constant areal target thickness of $\rho_\text{areal} = \unit[5 \times 10^{18}]{atoms/cm^2}$ are shown with a solid line assuming a jet diameter of $\unit[1]{mm}$. The points of intersection between the curves and the solid line represent possible working stagnation conditions. Due to the cluster formation process and the resulting advantages the lowest temperature and pressure combination should be chosen. }
    \label{fig:flow}   
\end{figure}

\paragraph{Design Stagnation Conditions}
Cluster-jet targets provide high and constant target thicknesses, e.g. directly behind the nozzle, which are determined by both the stagnation conditions of the gas in front of the Laval nozzle and the nozzles geometry. Laval nozzles have a specific convergent short inlet zone which merges after the narrowest inner diameter in a long divergent outlet zone. Assuming a homogeneous distribution within the jet beam, the volume density is defined by the mass flow $\dot m$ through the nozzle, the velocity $v$ of the clusters, the cross section $A_\text{beam}$ of the cluster beam, the Avogadro constant $N_A$ and the molar mass $M$ of the gas:  
	\begin{equation}
	\rho_\text{volume}=\frac{\dot m}{v \cdot A_\text{beam}}\cdot \frac{N_A}{M}.
	\label{eq:rhovolume}
	\end{equation}
The mass flow $\dot m$ itself is directly proportional to the volume flow $q_v$:
	\begin{equation}
	\dot m=\frac{q_v \cdot M \cdot p_\text{N}}{R \cdot T_\text{N}},
	\end{equation}
wherein $p_\text{N} = \unit[1.01325]{bar} $ and $T_\text{N} = \unit[273.15]{K} $ are the normal pressure and normal temperature, and $R$ the universal gas constant. The volume flow in turn is determined by the stagnation conditions of the gas in front of the nozzle $p_0$ and $T_0$, the cross-section area of the nozzle at the narrowest point $A^*$ and the heat capacity ratio $\kappa=C_p/C_V$ \cite{Wutz1988}:   
	\begin{equation}
	q_V = A^* \frac{p_0}{\sqrt{M T_0}} \frac{T_\text{N}}{p_\text{N}} \left( \frac{2}{\kappa + 1} \right)^{\frac{\kappa +1}{2 \left( \kappa -1 \right)}} \sqrt{\kappa R}.
	\label{eq:volumeflow}	
	\end{equation}
Consequently, the volume flow through the nozzle and the areal target thickness can be determined in dependence of the stagnation conditions of the gas in front of the nozzle. Here hydrogen is used as the first desired target material (see Figure \ref{fig:flow}). Furthermore, the volume density $\rho_\text{volume}$ can be converted to the areal target thickness $\rho_\text{areal}$ by an integration along the target beam diameter \cite{PhdKoehler}:
	\begin{equation}
	\rho_\text{areal} = 2 \int\limits_{0}^{\infty} \rho_\text{volume} \left( r \right) dr.
	\label{eq:arealthickness}
	\end{equation}

The calculations shown in Fig. \ref{fig:flow} were made for a Laval nozzle with a narrowest inner diameter of $d^* = \unit[0.5]{mm}$ and an outlet diameter of $d = \unit[1]{mm}$. In detail graphs of same volume flows feasible with different pressure and temperature combinations are shown. Additionally, a constant and in the range of the desired areal target thickness of $\rho_\text{areal} = \unit[5 \times 10^{18}]{atoms/cm^2}$ is assumed and plotted for all possible stagnation conditions. In order to reach the designed target thickness, various pressure and temperature combinations are feasible (blue dots). It should be stressed, however, that at same thickness an increase in temperature and pressure leads to an increase of the nozzle flow. To improve the vacuum conditions in the scattering chamber and also in the accelerator beam line, which results in minimized background signals, the lowest achievable temperature and pressure pair to realise the target thickness should be chosen as operation conditions of the jet target. Additionally, a decrease in temperatures leads to an onset and increase of the cluster formation process. The resulting cluster beam has the advantage, that it is well defined and less divergent in lateral direction compared to a conventional gas jet beam. Moreover, the target thickness does not decrease as rapidly with the distance from the nozzle. The underlying reason is that the mean size of the clusters increases with decreasing temperature and is predictable via the Hagena's scaling law \cite{Hagena_2}:
	\begin{equation}
	N = A_N \left( \frac{\gamma^*}{1000} \right) ^{\gamma_N},
	\end{equation}
wherein $N$ is the average cluster size, $\gamma^*$ the Hagena parameter, $A_N = 33$ and $\gamma_N = 2.35$ are empirical values for $\gamma^* > 1800$ \cite{Hagena_1, Hagena_2}. The Hagena parameter $\gamma^*$ itself is given by 
	\begin{equation}
	\gamma^* = \frac{\hat{k} p_0 \left( \frac{0.74 d_n}{\tan \alpha_{1/2}} \right)^{0.85}} {T_0^{2.29}},
	\end{equation}
with $\hat{k}_{H_2} = 184$ being a gas dependent constant \cite{k_hydrogen}, $p_0$ the stagnation pressure and $T_0$ the temperature, $d_n$ the nozzle diameter, and $\alpha_{1/2}$ the expansion half angle of the nozzle outlet. Successfully performed measurements of cluster mass distributions at the University M\"unster demonstrated, that the Hagena's scaling law agrees well to the results obtained with hydrogen cluster beams except for a constant underestimation factor of $2.6$ \cite{PhdKoehler}. This factor might be caused by the fact that the Hagena formula was mostly investigated for rare gases and not for hydrogen.\\

\begin{figure}[!ht]
    \centering
    \includegraphics[trim = 0mm 0mm 0mm 0mm, clip, width=0.4\textwidth]{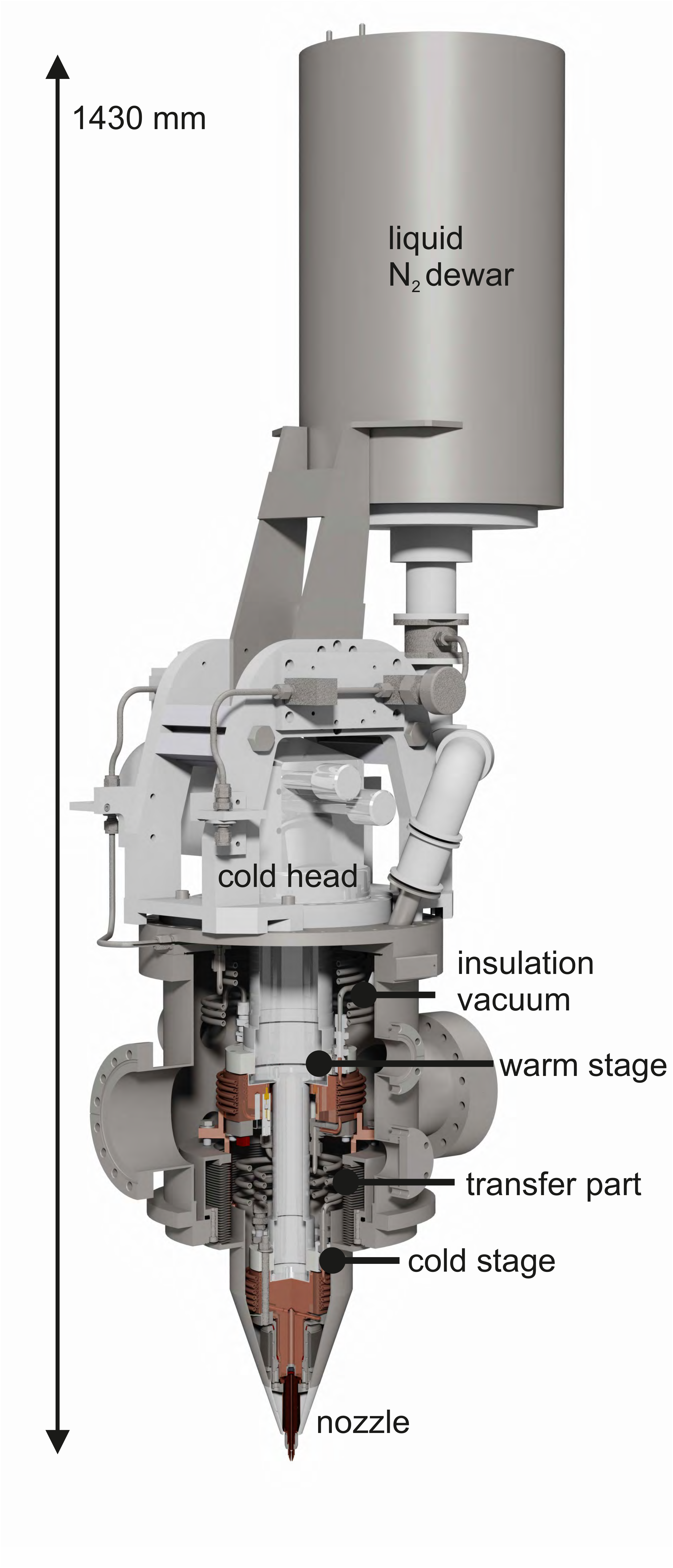}
    \caption{CAD drawing of the jet target for MAMI and MAGIX. The liquid nitrogen dewar in combination with the cold head is used to cool the gas down to temperatures around $T = \unit[40]{K}$ if high gas flows are required. The nozzle is mounted with a nozzle extension on the cold stage of the cold head. Additionally, to achieve a high acceptance for the electron spectrometers, the jet target has a conical tip with $\unit[20]{^\circ}$. Generated by D. Bonaventura, WWU M\"unster.}
    \label{fig:TargetCAD}   
\end{figure}

For the MAGIX target, two different Laval nozzle designs for the jet beam are chosen. The first Laval nozzle has a narrowest inner diameter of $d^* = \unit[0.5]{mm}$ and an outlet diameter of $d = \unit[1]{mm}$ (cf. Fig. \ref{fig:flow}) and the second one the same narrowest inner diameter, but an outlet diameter of $d = \unit[2]{mm}$. For both nozzles a temperature of $T = \unit[40]{K}$ and a volume flow of $q_v = \unit[40]{\ell/min}$ are determined as design operating stagnation conditions. This results for the first nozzle (outlet diameter $d = \unit[1]{mm}$) in an areal target thickness of $\rho_\text{areal} = \unit[5 \times 10^{18}]{atoms/cm^2}$ and for the second nozzle (outlet diameter $d = \unit[2]{mm}$) in an areal target thickness of $\rho_\text{areal} = \unit[2.4 \times 10^{18}]{atoms/cm^2}$. By this, depending on the experimental needs by choosing an appropriate nozzle temperature, the target can be used as conventional gas jet target or as cluster-jet target. Moreover, the feasibility of the target to be operated at such low temperatures will enable to produce also cluster beams from heavier gases such as Ar, O$_2$, N$_2$, Xe, and many more.\\

\begin{figure}[!ht]
    \centering
    \includegraphics[trim = 0mm 150mm 105mm 0mm, clip, width=0.47\textwidth]{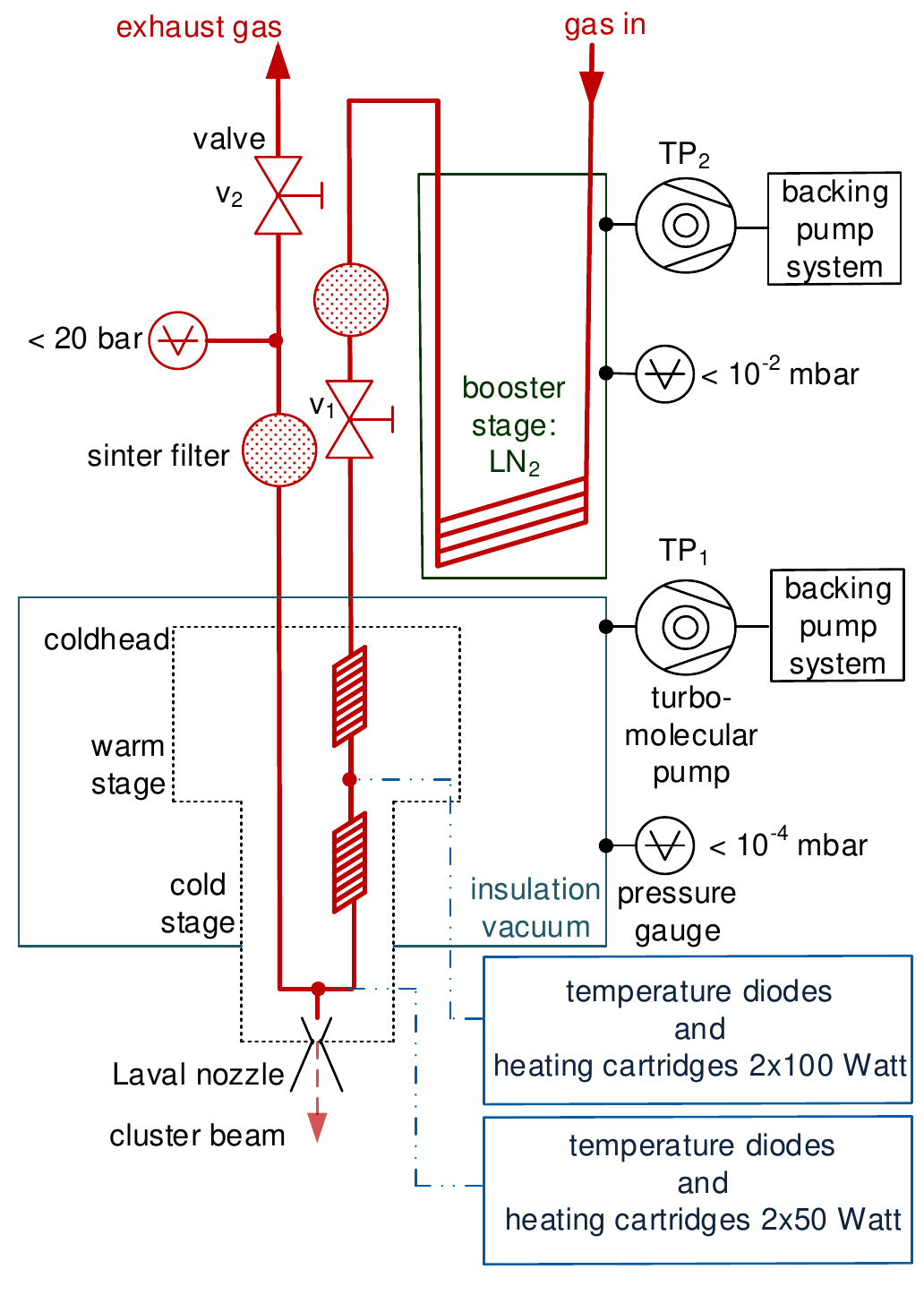}
    \caption{Schematic view of the gas (red) and vacuum system of the jet target for MAMI and MAGIX. The gas is precooled in the booster stage and is then directed to the cold head and the nozzle. The double-walled walls of the booster stage and the insulation vacuum chamber are pumped via turbomolecular pumps to minimize the heat exchange. Additionally, the two stages of the cold head are equipped with temperature diodes and heat cartridges to monitor and control the temperature of the used gas (cf. table \ref{tab:targetdevices}).}
    \label{fig:TargetVisio}   
\end{figure}

\paragraph{Design of the Target}
Based on the described requirements, e.g. the target thickness, and the design operating stagnation conditions, the jet target for MAGIX was designed at the University of M\"unster (CAD drawing see Fig. \ref{fig:TargetCAD}, gas and vacuum system see Fig. \ref{fig:TargetVisio}). 
To achieve the cooling at a volume flow of $q_v = \unit[40]{\ell/min}$ hydrogen down to temperatures of around $T = \unit[40]{K}$, a two stage cold head is used (see table \ref{tab:targetdevices}). This powerful cold head device allows for lowest temperatures of $ \unit[13]{K}$ without hydrogen gas flow, at low hydrogen gas flows, i.e., $T = \unit[20]{K}$ at $q_v = \unit[5.5]{\ell/min}$, and even at relatively high hydrogen gas flows of $q_v = \unit[25]{\ell/min}$ temperatures of $T = \unit[50]{K}$ can be realised. The hydrogen gas is directed through pipes around the two cooling stages (warm and cold stage) of the cold head. At these two stages the pipes are made out of copper to ensure an optimal heat transfer between the cold head and the used gas. The transfer parts, i.e., the section between the two stages and the pipes towards the gas inlet section, are made out of stainless steel, to guarantee a minimal heat exchange of the stages operated at different temperatures. Additionally, the cold head is enclosed in an insulation vacuum chamber, pumped via a turbomolecular pump (see figure \ref{fig:TargetVisio}), to insulate the cooled parts. Due to the use of bellows the insulation vacuum reaches up to the nozzle area. Therefore, only a small amount of the cooled device (nozzle area) has direct contact to the vacuum in the scattering chamber which ensures minimal heat transfer. Thus, the target reaches lowest temperatures due to minimized transfer even at high gas flows through the nozzle which corresponds to higher pressures in the interaction vacuum chamber. \\
To offer the possibility to realise even lower temperatures, a booster stage is installed before the gas enters the cold head cooling system. This pre-cooling system is filled with $ \unit[14]{\ell}$ of liquid nitrogen and the hydrogen gas is directed through windings of copper pipes inside the vessel, whose walls are double-walled with vacuum in between to minimize the nitrogen consumption due to heat losses. The pre-cooled gas then enters the cold head system through an insulated pipe via a vacuum feedthrough. The efficient cold head in combination with the pre-cooling system of the target, makes it very well suited for a wide range of application. Without any change in the setup of the target, the target can operate as a gas jet target at normal temperature, but otherwise cool down very high gas flows of $q_v = \unit[40]{\ell/min}$ down to temperatures of $T = \unit[40]{K}$, and lower gas flows to even lower temperatures, to realise a cluster beam.\\
The nozzle used to produce the cluster beam is assembled with a nozzle extension on the cold stage of the cold head for a maximal detector acceptance. This extension can easily be adapted to the experimental needs and allows for a variation of the nozzle position. Furthermore, this setup allows for a simple exchange of the nozzle, which is essential for the different experimental issues due to the varying gas types and target thicknesses realisable with different nozzle geometries. The nozzle is sealed with a special polyimide sealing, which supports the simplicity of a nozzle exchange. The sealing is reusable and in comparison to the previous used indium sealings, the new one ensures a clean extraction, a non tilting mounting of the nozzle, it avoids the necessary cleaning of the nozzle and devices from the indium, and prevents from fine particles in the gas system which might block the nozzle. Moreover, the sealing is ideally suited for this application due to its cold resistance to ensure the tightness between nozzle and nozzle seat. To ensure a blocking of the fine nozzle due to impurities in the gas, two sinter filters are installed in the pipe. The special polyimide sealing for the nozzle sealing is also used inside the target for sealing the individual components due to the radiation resistance needed for the application at an electron accelerator. In order to allow for a high detector acceptance, the target has a conical tip of $\unit[20]{^\circ}$. Furthermore, to monitor and control the pressure and temperature of the gas in front of the nozzle, a baratron, temperature diodes, and heat cartridges are installed in the two stages of the cold head (details see table \ref{tab:targetdevices}). The jet target allows for a vacuum-tight flexible connection via a bellow to the future scattering chamber, which enables the possibility to install a system to adjust the target in regard to the accelerator beam.\\
The design of the target is sophisticated and optimized to the requirements on current and future experiments using a jet target, which find their applications in various fields of research, like their use in combination with hadron, lepton or laser beams. Therefore, this target served already as the basis for forthcoming targets, i.e., the CryoFlash target which is currently built up at the University of M\"unster. Furthermore, the compact design of the target enables for an easy and fast installation or removal at the experimental facilities and is well suited for complex structures, because it takes low physical space and can be installed in any orientation (with possible rotation of the N$_2$ dewar). The different working conditions (gas jet target or cluster-jet target) in combination with the possibility to use diverse gases as target material, and the simple exchange of the nozzle to realise various nozzle geometries, makes this target to a powerful state-of-the-art target.   

\begin{table}[ht]
\caption{Specifications of the assembled devices in the target to cool the gas and to monitor and control the gas temperature of the target.}
	\centering
		\begin{tabular}{c|c}
		device  & specification  \\ 
	 	\hline	 			
		cold head  & Leybold Coolpower 10MD \\
							& 1. Stage: $\unit[110]{W}$ at $\unit[80]{K}$ \\
							& 2. Stage: $\unit[18]{W}$ at $\unit[20]{K}$  \\
		compressor & Leybold Coolpak 6000H\\
		temperature control & Lakeshore 336    \\
		temperature diodes & 1. Stage: $2 \times $ DT 670 (Lakeshore) \\
						   & 2. Stage: $2 \times $ DT 670 (Lakeshore)   \\
		heating cartridges   & 1. Stage: $2 \times \unit[100]{W}$ (Lakeshore) \\
						   & 2. Stage: $2 \times \unit[50]{W}$ (Lakeshore)
		
		\end{tabular}
		\label{tab:targetdevices}
\end{table}

\section{Results}
\begin{figure}[!ht]
    \centering
    \includegraphics[trim = 0mm 182mm 145mm 0mm, clip, width=0.4\textwidth]{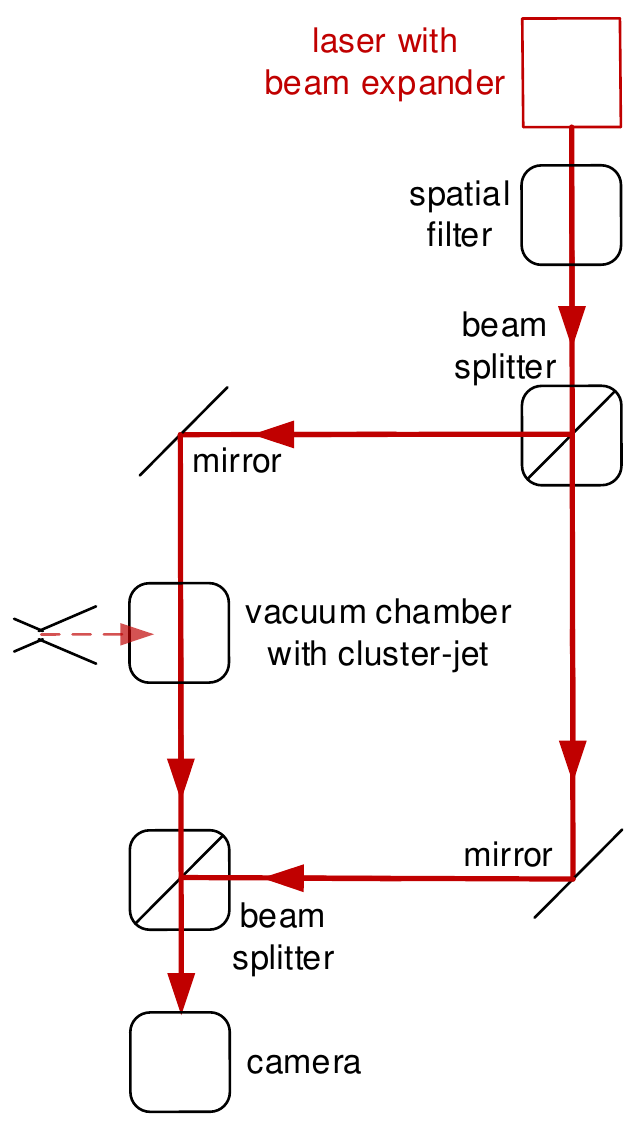}
    \caption{Sketch of the Mach Zehnder interferometer used to perform jet beam studies with the target for MAMI and MAGIX. A laser beam is widened and adjusted with a beam expander and a spatial filter. With a beam splitter the beam is divided into two sub-beams. The reference beam is directed via optical components to a camera. The analysis beam is guided with a mirror through a vacuum chamber, where the cluster beam is located and interferes then with the reference beam on the camera sensor. }
    \label{fig:MZVisio}   
\end{figure}

In autumn 2016, the jet target for MAMI and MAGIX was built up and set successfully into operation at the University of M\"unster. It was possible to successfully demonstrate that the target achieves the design operation stagnation conditions at given hydrogen gas flows. Furthermore, to perform detailed cluster beam analyses, a Mach Zehnder interferometer was assembled (Figure \ref{fig:MZVisio}) \cite{PhdGrieser}. 

\begin{figure}[!ht]
    \centering
    \includegraphics[trim = 0mm 0mm 0mm 0mm, clip, width=0.48\textwidth]{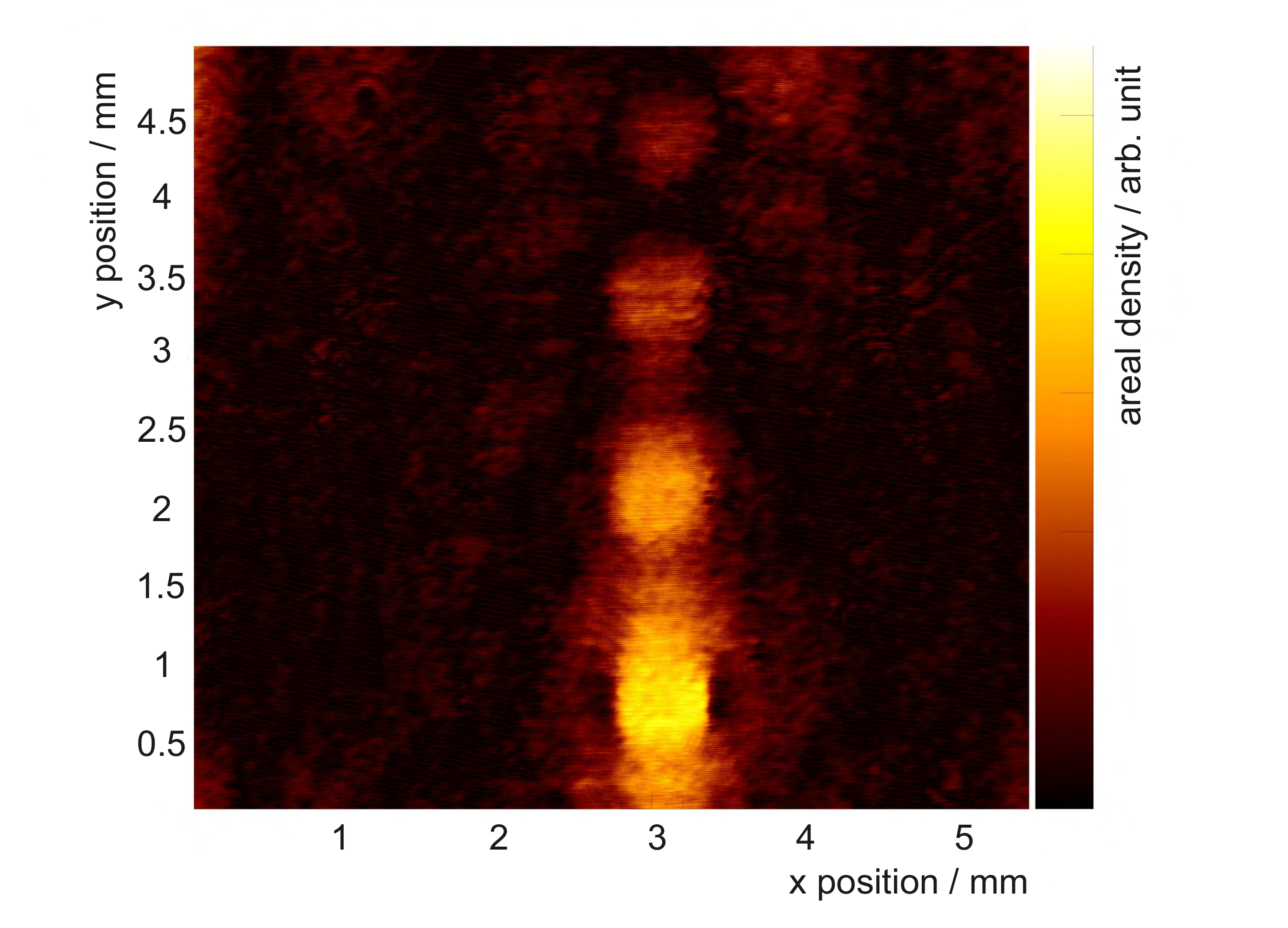}
    \caption{Expanding nitrogen jet beam ($p_0 = \unit[20]{bar}$, $ T_0 = \unit[288]{K}$) from bottom to top into atmospheric pressure recorded with a Mach Zehnder interferometer. The nozzle (not visible) with a narrowest inner diameter of $d^* = \unit[0.5]{mm}$ and an outlet diameter of $d = \unit[2]{mm}$ is located between $ \unit[2]{mm}$ - $ \unit[4]{mm}$ in $x$ position at the bottom directly under the edge of the visible area. A jet beam with its typically node structure over a distance of more than $ \unit[4.5]{mm}$ can be observed.}
    \label{fig:gasjetair}   
\end{figure}

\begin{figure}[!ht]
    \centering
    \includegraphics[trim = 0mm 40mm 30mm 0mm, clip, width=0.48\textwidth]{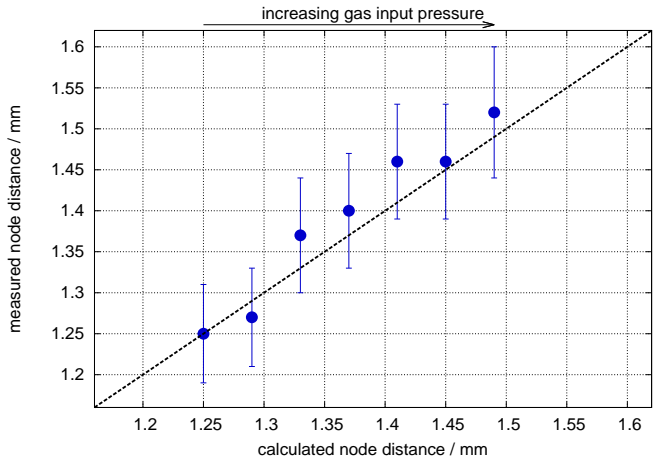}
    \caption{Comparison between the empirical description of the distance between the nodes via Eq. \ref{eq:nodes} and the measurements taken at gas input pressures between $p_0 = \unit[14]{bar}$ and $p_0 = \unit[20]{bar}$ in steps of $\unit[1]{bar}$ (cf. Figure \ref{fig:gasjetair}). A higher pressure results to a larger distance between the nodes.}
    \label{fig:nodes}   
\end{figure}

\begin{figure}[!ht]
    \centering
    \includegraphics[trim = 0mm 0mm 0mm 0mm, clip, width=0.48\textwidth]{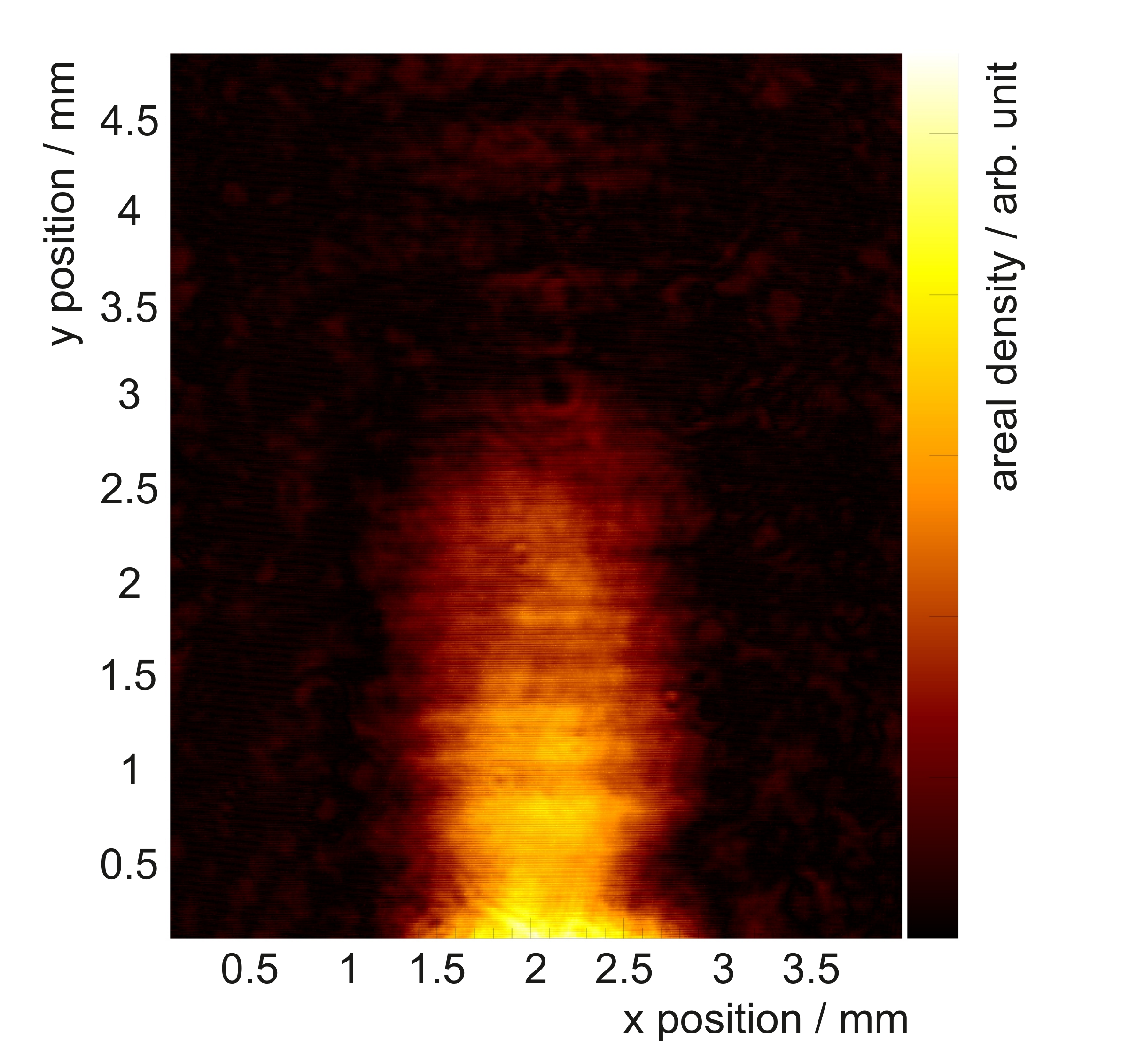}
    \caption{Expanding jet beam from bottom to top into vacuum ($p_a = \unit[4]{mbar}$) recorded with a Mach Zehnder interferometer. The nozzle, the gas, and the pressure and temperature of the gas in front of the nozzle are equal as in Figure \ref{fig:gasjetair}. The nozzle (not visible) is located, but not visible at the bottom between $ \unit[1]{mm}$ - $ \unit[3]{mm}$. A jet beam can be observed, which expands rapidly in lateral direction and vanishes already after $ \unit[3]{mm}$.}
    \label{fig:gasjetvacuum}   
\end{figure}

\begin{figure}[!ht]
    \centering
    \includegraphics[trim = 0mm 0mm 0mm 0mm, clip, width=0.3\textwidth]{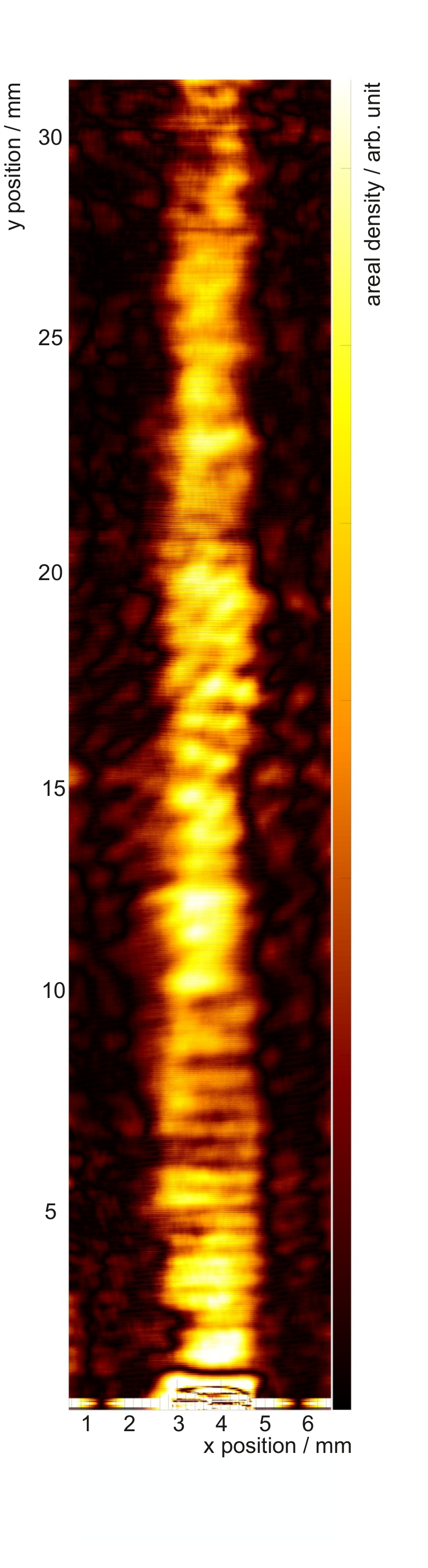}
    \caption{A hydrogen jet beam recorded at the design stagnation conditions for MAGIX ($T = \unit[40]{K}$ and $q_v = \unit[40]{\ell/min} $) with a Mach Zehnder interferometer. The nozzle tip is visible as a bright area between $ \unit[2.5]{mm}$ - $ \unit[4.5]{mm}$ at the bottom with a hight of around $ \unit[0.5]{mm}$. A well defined jet beam (direction from bottom to top) leaves the nozzle and is clearly visible over the full field of view, i.e. $ \geq \unit[30]{mm}$ from the nozzle. The colour code represents the relative target thickness in a linear scale.}
    \label{fig:clusterbeam}   
\end{figure}

The interferometer offers the possibility to investigate the absolute and relative target thickness within the cluster beam and additionally to calculate the shape and the range of the beam. Such studies are of great relevance in order to design an effective beam catcher system installed close to the interaction zone in order to improve the vacuum conditions in the scattering chamber. \\
The interference of the reference laser beam and the analysis beam, which is directed through a vacuum chamber where the cluster beam is located, is monitored by a camera sensor. The superposition of the two beams generates typically interference bands, and the nozzle is visible as a dark shadow. A phase shift in the interference bands corresponds to a change in the refractive index caused by a variation in the target thickness. Since the target will also be used for experiments with heavier gases, first measurements were performed with nitrogen instead of hydrogen as target material. The used Laval nozzle had a narrowest inner diameter of $d^* = \unit[0.5]{mm}$ and an outlet diameter of $d = \unit[2]{mm}$ and was produced at the University of M\"unster by an electroplating process. This nozzle production process enables the possibility to manufacture nozzles with narrowest inner diameter, e.g. between $ \unit[0.012]{mm}$ - $ \unit[0.5]{mm}$, and variable inlet and outlet geometry. The nitrogen gas expands at $ T_0 = \unit[288]{K}$ with a pressure of $p = \unit[20]{bar}$ into the non-evacuated vacuum chamber, so that an expansion into atmospheric pressure ($p_a = \unit[1.013]{mbar} $) can be observed. Figure \ref{fig:gasjetair} shows the jet beam expanding from bottom to top after reconstruction of the target thickness from the interference pictures. For details on the applied method, using the fast forward Fourier transformation to calculate the phase shifts which are proportional to the target thickness, the reader is refered to \cite{MALessman}. The colour code in Figs. \ref{fig:gasjetair} - \ref{fig:clusterbeam} represents the relative target thickness in a linear scale. 

Clearly visible is the typical node structure with maximum and minimum target thicknesses. The jet beam is reflected at the free-jet boundary conditions and compressed to a minimal cross-sectional area, the so called Mach disk shock \cite{Pauly}. Additionally, typically for a gas jet beam the target thickness decreases with increasing distance from the nozzle. The node structure can be described well by \cite{Pauly}
	\begin{equation}
	\frac{x_M}{d^*} = 0.67 \cdot \sqrt{\frac{p_0}{p_a}}
	\label{eq:nodes}
	\end{equation}
wherein $x_M$ is the distance between the thickness nodes and $p_a$ the pressure in the vacuum chamber. Due to the ambient pressure of $p_a = \unit[1.013]{mbar}$ the first nodes are located within the nozzle and not visible in the picture taken with the Mach Zehnder interferometer. Because of the friction between the jet beam and the residual gas, the distances between the subsequent nodes decreases with increasing distance from the nozzle. As a good approximation, the distance between the first two visible nodes after the nozzle exit can be estimated and compared with the empirical description (Figure \ref{fig:nodes}). These measurements were taken at gas input pressures between $p_0 = \unit[14]{bar}$ and $p_0 = \unit[20]{bar}$ in steps of $\unit[1]{bar}$ and are in excellent agreement with the prediction of Eq. \ref{eq:nodes}. Consequentially, the assembled Mach Zehnder interferometer is ideally suited to study the shape and the thickness structures of the jet beam, which is from highest interest to optimize an optional catcher system for the jet beam after the interaction with the accelerator beam.\\

The jet beam shape changes significantly if the vacuum chamber is evacuated. In a following investigation the same nozzle, same gas, and pressure and temperature of the gas in front of the nozzle were used. Only the pressure in the chamber was reduced to $p_a = \unit[4]{mbar}$ by evacuating the chamber with a roots pumping system. The resulting reconstructed thickness distribution is presented in Figure \ref{fig:gasjetvacuum} and shows a jet beam which expands rapidly in lateral direction. After a distance of already $ \unit[3]{mm}$, the jet beam has no appreciable thickness and disappears in the background vacuum.     

Such a jet beam is not feasible for an experiment like MAGIX due to the high gas load remaining in the scattering chamber increasing the background rates. Instead, a beam catcher system has to be used which requires a well defined beam with a large range in order to leave the scattering chamber to large extend. Therefore, the gas in front of the nozzle will be cooled down to temperatures of $T = \unit[40]{K}$, where the cluster formation process takes place. Measurements with the Mach Zehnder interferometer at the design stagnation conditions ($T = \unit[40]{K}$ and $q_v = \unit[40]{\ell/min} $) and hydrogen as target material were performed.\\
Figure \ref{fig:clusterbeam} shows the reconstructed thickness obtained at this measurement. The nozzle is visible as a bright shadow at the bottom with a diameter of $\unit[2]{mm}$ and the cluster beam direction is from bottom to top. Different to the observations using a jet beam at room temperature the cluster beam is well defined over a long distance ($> \unit[30]{mm}$) and has a diameter of roughly $\unit[2]{mm} - \unit[3]{mm}$. In comparison to the empirical description of the node structure (cf. Eq. \ref{eq:nodes}), in distances of $ \unit[7.6]{mm}$ nodes should occur, but due to the formation of massive clusters no clear distinct thickness nodes are visible within the jet beam. \\
Although the Mach Zehnder device allows for the determination of absolute target beam thicknesses, here only relative values are presented. Instead, the absolute scale can be determined with much higher precision by using Eqs. \ref{eq:rhovolume} - \ref{eq:arealthickness} in combination with the well known nozzle exit diameter. Furthermore, the Mach Zehnder studies demonstrated the reproducibility and stability of the target. The temperature and the pressure of the gas in front of the nozzle, and the gas flow were found to be stable during the full time with accuracies of $\Delta T < \unit[1]{K}$, $\Delta p < \unit[0.1]{bar}$, and $\Delta q_v < \unit[1]{\ell/min}$ given by the readout resolution of the target control devices. Consequently, the target thickness, the beam shape, and the position of the target beam remains the same, so that there is no need for a re-adjustment of the jet beam during future beam times.\\
In the present measurement of Fig. \ref{fig:clusterbeam}, an areal target beam thickness of $\rho_\text{areal} = \unit[2.4 \times 10^{18}]{atoms/cm^2}$ in the center of the nozzle exit has been calculated. Obviously an operation in the cluster mode allows for the possibility to provide a well defined target beam, which can be extracted after the interaction zone by an appropriate beam catcher device with an opening aperture matching the target beam diameter. Due to the absence of a beam catcher resulting in a relatively high gas load to the examination chamber equipped with vacuum pumps of only limited pumping speed of $\sim \unit[400]{m^3/h} $, the chamber pressure reached values of $\unit[6]{mbar}$. Later at the MAMI accelerator and at the MAGIX experiment the vacuum will be improved significantly by the use of a powerful vacuum pumping system, i.e. consisting of two turbomolecular pumps and a roots pumping system with pumping speeds of $ \unit[5000]{\ell/s}$ and $ \unit[7000]{m^3/h}$, respectively. Upcoming studies will lead to an optimized beam catcher system to significantly improve the vacuum at the interaction point \cite{PhdAulenbacher}. 

\section*{Summary}
A high-performance state-of-the-art target was designed, built up and set successfully into operation at the University of M\"unster considering the requirements of the experimental setup of MAGIX. The target allows for different working stagnation conditions (temperature and gas flow through the Laval nozzle) in combination with different nozzle geometries and different gases as target material. Therefore, it is highly suited for the MAGIX experiment and serves as a basis for future targets for forthcoming experiments, e.g. CryoFlash. In this publication, it was shown, that the target fulfils exceedingly all requirements. Moreover, two suitable Laval nozzles manufactured for the design stagnation conditions ($T = \unit[40]{K}$ and $q_v = \unit[40]{\ell/min} $) for first measurements are discussed. These parameters allow for an areal target thickness of up to $\rho_\text{areal} = \unit[5 \times 10^{18}]{atoms/cm^2}$ directly behind the nozzle. \\
Measurements performed with a Mach Zehnder interferometer demonstrate the difference between a conventional gas jet with a rapid divergence in lateral direction and a well defined cluster beam with a high range of the thickness by cooling the gas in front of the Laval nozzle down to temperatures of $T = \unit[40]{K}$.\\
The stable and high performance of the target was shown in first successful commissioning beam times at the A1 hall at the MAinzer MIkrotron (MAMI) in summer 2017 and first cross sections of the elastic electron-proton scattering with excellent low background were taken (for more details see \cite{PhdGrieser, MABrand}). Moreover, the quality of these data proved the capability of the new experimental setup using the jet target which is ideally suited for precision measurements. Furthermore, the A1 experiment offers the possibility to scan the jet beam of the target with the electron beam by deflection magnets. The resulting jet beam profiles give further information about the thickness and the shape of the jet beam, which are of high interest for both the jet beam optimization and the later data analysis (\cite{PhdGrieser, MABrand}).

\section*{Acknowledgements}
The authors would like to thank the other members of the MAGIX collaboration for the 
successful cooperation as well as the A1 collaboration and the MAMI crew for the great
support during the related experimental runs at MAMI.\\
The work provided by the teams of our mechanical and electronic workshops is very much appreciated and we thank them for the excellent manufacturing of the various components. \\
Furthermore, we would like to thank R. Balske and D. Veith for their great support and preliminary studies.


\bibliography{references}

\end{document}